\documentclass[twocolumn,showpacs,preprintnumbers,amsmath,amssymb,letterpaper]{revtex4}%

\usepackage{graphicx}
\usepackage{dcolumn}
\usepackage{bm}
\usepackage{epsf}
\usepackage{hyperref}

\def\tR{{\tilde R}}
\def\tF{{\widetilde F}}

\def\cU{{\cal U}}

\def\cW{{\cal W}}

\def\s{{\sigma}}

\begin{document}
\preprint{}
\title{Exact Quantum State of Collapse and Black Hole Radiation \\}
\author{Cenalo Vaz${}^{a\dagger}$, L. Witten${}^{a\ddagger}$ and T.P. Singh${}^{b}$}
\affiliation{$^{a}$Department of Physics,\\
University of Cincinnati, Cincinnati, OH 45221-0011, USA.\\
$^{\dagger}$\textrm{Email address:
\href{mailto:vaz@physics.uc.edu}{\texttt{vaz@physics.uc.edu}}}\\
$^{\ddagger}$\textrm{Email address:
\href{mailto:witten@physics.uc.edu}{\texttt{witten@physics.uc.edu}}}\\}
\affiliation{$^{b}$Tata Institute of Fundamental Research,\\
Homi Bhabha Road, Mumbai 400 005, India.\\
\textrm{Email address:
\href{mailto:tpsingh@nagaum.tifr.res.in}{\texttt{tpsingh@nagaum.tifr.res.in}}}}

\begin{abstract}
\noindent We construct an exact quantum gravitational state describing the
collapse of an inhomogeneous spherical dust cloud using a lattice
regularization of the Wheeler-DeWitt equation. In the semiclassical
approximation around a black hole, this state describes Hawking radiation. We
show that the leading quantum gravitational correction to Hawking radiation
renders the spectrum non-thermal.

\end{abstract}

\maketitle

\section{\label{sec:intro}Introduction}

The end state of the gravitational collapse of a compact object is an
important problem for which a full understanding has not yet been achieved. It
is not completely clear what changes a quantum theory of gravity will bring
about in our understanding of Hawking radiation, the entropy of a black hole,
the information loss paradox and the nature of the gravitational singularity.
This is because it has not been easy to develop dynamical models of quantum
gravitational collapse in the various candidate theories of quantum gravity.

Over the last few years, useful and interesting developments in formulating
the quantum gravitational collapse problem have taken place, within the
framework of canonical quantum general relativity. Kastrup and Thiemann
\cite{kt} and Kucha\v r \cite{kuc1,kuc2,kuc3} have pioneered the technique of
midisuperspace quantization, a canonical quantization of a gravitational
system with symmetries, and yet possessing infinitely many degrees of freedom.
A prime example of this is the canonical quantization of the Schwarzschild
geometry associated with an eternal black hole \cite{kuc2}.

We have applied these earlier works to develop a canonical description
\cite{vws} of the collapse of a spherical time-like dust cloud, which is
described by the LeMa\^\i tre-Tolman-Bondi family of metrics \cite{ltb1}.
Quantization of the classical dust system leads to the Wheeler-DeWitt equation
for the wave-functional describing the quantum collapse. The quantum Schwarzschild
black hole then becomes a special case of our model. We have successfully
employed this equation to derive the
Bekenstein mass spectrum \cite{bek1} and statistical entropy of the charged
and uncharged black hole \cite{cvlw1} and, in a recent work \cite{vksw}, we
showed that the semiclassical (WKB) treatment of the Schwarzschild black hole
in this canonical picture describes Hawking radiation. The midisuperspace
program has also been used to describe the collapse of null dust, with
interesting conclusions \cite{lwf,biku,haki,ha,vws2}. All these works
demonstrate the overall consistency of the program.

Apart from the usual operator ordering ambiguities, a problem with the
midisuperspace quantization program is that the Wheeler-DeWitt equation
involves two functional derivatives taken at the same point and requires
regularization in all but the effectively minisuperspace models. Any attempt
at regularization, however, leads to ambiguities that are not easy to resolve.
Our purpose, in the present article, is to address this problem by applying a
lattice regularization scheme to the canonical quantization of LeMa\^\i
tre-Tolman-Bondi (LTB) models.

We first construct a stationary state solution of the diffeomorphism
constraint. Upon discretization, the wave-functional reduces to a tensor
product over shell wave-functions, one at each lattice point. The
Wheeler-DeWitt equation is well defined on the lattice as an infinite set of
Schroedinger equations, one for each shell. The results are independent of
how the lattice is discretized when the size of all lattice cells approaches
zero. Even though the regularization is on a lattice, diffeomorphism invariance
is recovered when the lattice spacing approaches zero.
The shell wave-functions reproduce Hawking's thermal radiation near the
horizon in the WKB limit and the first quantum gravitational correction to
Hawking's formula renders the spectrum non-thermal. We believe that
our construction opens up a useful avenue for investigating some of the
important unresolved issues related to black hole physics mentioned above.

The paper is organized as follows: in section \ref{sec:ltb} we review our
pervious quantization of the LeMa\^\i tre-Tolman-Bondi collapse models. In
section \ref{sec:lattice} we construct a lattice to regularize the
Wheeler-DeWitt equation and discuss its properties. Solutions of the
Wheeler-DeWitt equation are discussed in section \ref{sec:solutions} and
Hawking radiation is reexamined in section \ref{sec:hawking}. We conclude with
a few comments in section \ref{sec:discussion}

\section{\label{sec:ltb} Canonical Quantization of LTB Collapse}

The LeMa\^\i tre-Tolman-Bondi models \cite{ltb1} constitute a complete solution of the Einstein
equations for a matter continuum of inhomogeneous dust, {\it i.e.,} they are solutions of the
spherically symmetric Einstein's field equations, $G_{\mu\nu} = - 8\pi G T_{\mu\nu}$, with
vanishing cosmological constant and with stress-energy describing inhomogeneous, pressureless
dust given by $T_{\mu\nu} = \epsilon u_\mu u_\nu$. The solutions are determined by two
arbitrary functions, $F(r)$ and $E(r)$, of a spatial label coordinate, $r$, and given in
co-moving coordinates as
\begin{eqnarray}
ds^2 &=& d \tau^2 - \frac{R'^2}{1 + E} dr^2 - R^2 d\Omega^2,\cr
\epsilon &=& \frac{\tF}{R^2 \tR},~~ R^* = \pm \sqrt{E + \frac{F}{R}},
\label{ltbsol}
\end{eqnarray}
(we have set $8\pi G = 1 = c$) where $R(\tau,r)$ is the physical radius of a spherical
shell labeled by $r$. A prime represents a derivative w.r.t. $r$ and a star (${}^*$) represents
a derivative with respect to the dust proper time, $\tau$. The arbitrary functions, $E(r)$
and $F(r)$, are interpreted respectively as the energy and mass functions. The energy
density of the collapsing matter is $\epsilon(\tau,r)$, and the negative sign in the third
equation above is required to describe a collapsing cloud. Its general solution is given up
to an arbitrary function $\psi(r)$ of the shell label coordinate. This arbitrariness reflects
only a freedom in our choice of units {\it i.e.,} at any given time, say $\tau_o$, the function
$R(\tau_o,r)$ can be chosen to be an arbitrary function of $r$.

The mass function, $F(r)$, represents the weighted mass (weighted by the factor $\sqrt{1+E}$)
contained within the matter shell labeled by $r$. If a scaling is chosen so that the physical
radius coincides with the shell label coordinate, $r$, at $\tau=0$, then it can be expressed
in terms of the energy density at $\tau=0$ according to
\begin{equation}
F(r) = \int \epsilon(0,r) r^2 dr,
\end{equation}
while the energy function, $E(r)$, can be expressed in terms of the initial velocity profile,
$v(r) = R^*(0,r)$, according to
\begin{equation}
E(r) = v^2(r) - \frac{1}{r} \int \epsilon(0,r) r^2 dr,
\end{equation}
showing that it represents the total energy (gravitational plus potential) of the matter-gravity
system. The marginally bound models, which we will consider in this paper, are defined by $E(r)=0$.
For the scaling referred to above, we must choose $\psi(r) = r^{\frac{3}{2}}$, whence the solution
of (\ref{ltbsol}) can be written as
\begin{equation}
R^{\frac{3}{2}}(\tau,r) = r^{\frac{3}{2}} - {\frac{3}{2}} \sqrt{F(r)}\tau.
\label{ltbrad}
\end{equation}
The epoch $R=0$ describes a physical singularity, whose singularity curve
\begin{equation}
\tau(r) = \frac{2r^{\frac{3}{2}}}{3\sqrt{F(r)}},
\end{equation}
gives the proper time when successive shells meet the central physical singularity.
Various models are obtained from choices of the mass function, $F(r)$. For example, the
Schwarzschild black hole is the marginally bound solution with $F(r) = 2M$, a constant.

Reinserting the constants $c,\hbar$ and $G$, and defining the dimensionless variable
$x=r/l_p$ (where $l_p=\sqrt{\hbar G/c^3}$ is the Planck length) and the dimensionless
mass function $f(x)=F/l_p$, the wave-functional of the quantized Einstein-Dust system
can be shown \cite{vws} to satisfy the Wheeler-DeWitt equation,
\begin{equation}
\left[  \frac{1}{c^{2}}\frac{\delta^{2}}{\delta\tau^{2}(x)}\pm\frac{\delta^{2}%
}{\delta R_{\ast}^{2}(x)}\pm\frac{f^{\prime2}}{4l_{p}^{2}|\mathcal{F}|}\right]
\Psi\lbrack\tau,R,f]=0, \label{wd}%
\end{equation}
where the prime denotes a derivative with respect to $x$. The upper sign refers to the
region outside the horizon, $R>F$, and the
lower sign to the region inside, $R<F$. We have also used ${\mathcal{F}}=1-F/R$ and
a radius function
\begin{equation}
R_{\ast}=\pm\int\frac{dR}{\sqrt{\left\vert \mathcal{F}\right\vert }}~~~~ \in
(-\infty,+\infty),
\label{rst}%
\end{equation}
defined so that the DeWitt configuration space metric is
manifestly flat. Invariance under spatial diffeomorphisms is implemented by the
momentum constraint \cite{vws},
\begin{equation}
\left[\tau^{\prime}\frac{\delta}{\delta\tau}+R_{\ast}^{\prime}\frac{\delta
}{\delta R_{\ast}}+f^{\prime}\frac{\delta}{\delta f}\right]  \Psi\lbrack
\tau,R,f]=0. \label{mc}%
\end{equation}
To complete the quantum theory, one must define an inner product on the
Hilbert space of wave-functionals. In \cite{vws}, we defined it in a natural
way, by exploiting the fact that the DeWitt super-metric \cite{deW} is
manifestly flat in the configuration space $(\tau,R_{\ast})$, in terms of the
functional integral
\begin{equation}
\left\langle \Psi_{1}|\Psi_{2}\right\rangle =\int_{R_{\ast}(0)}^{\infty
}\mathcal{D}R_{\ast}\Psi_{1}^{\dagger}\Psi_{2}. \label{ip}%
\end{equation}
Equations (\ref{wd})--(\ref{ip}) clearly imply a specific choice, albeit a
natural one, of operator ordering, for which no justification deeper than the
fact that they were shown to reproduce the Bekenstein mass spectrum
\cite{bek1,cvlw1} of the eternal black hole and its Hawking radiation in the
WKB approximation \cite{vksw} can be given at this time. This defect is
unfortunately endemic to current theories of quantum gravity.

\textit{Any} functional that is a spatial scalar will obey the momentum
constraint. In particular the functional
\begin{widetext}
\begin{equation}
\Psi[\tau,R,f] = \exp \left[-\frac{ic}{2l_{p}} \int_0^\infty dx f'(x)
{\cal W}(\tau(x),R(x),f(x)) \right],
\label{ansatz}
\end{equation}
\end{widetext}
is a spatial scalar if $\cW(\tau,R,F)$ has no explicit dependence on $x$ (equivalently,
$r$) and we will use this as a solution ansatz. If we further require
(\ref{ansatz}) to represent a stationary state then, noting that $F^{\prime}(r)/2$ is
the proper energy density of the collapsing cloud, we choose
\begin{equation}
\mathcal{W}(\tau,R,f)=\tau+\mathcal{U}(R,f).
\label{ans}%
\end{equation}
In a classical collapse problem, $f(x)$ is generally monotonically increasing up to
the boundary of the star, which is specified by some label $x_b$. The exterior of the
star is then matched to the Schwarzschild exterior and $f(x)$ is taken to be constant
for $x>x_b$. While this is a reasonable condition to impose upon the mass
function on the classical level, it is too strong in the quantum theory. To account for
Hawking radiation, it is necessary for $f(x)$ to be a monotonically increasing function
of $x$ over the entire range of $x$, as we do not expect to have a sharp boundary
between the collapsing matter and its exterior owing to the evaporation process.

\section{\label{sec:lattice} Lattice Construction}

\bigskip We choose a lattice by dividing space into cells, the size of the $j^{th}$ cell
being $\s_{j}$, and we go to the limit $\s_{j}\rightarrow0,\forall j$. The wave functional
and all results are completely independent of the choice of the $\s_{j}$. We define the
integral in the exponent of (\ref{ansatz}) as the limit of a Riemann sum over the
lattice points as the spacing goes to zero.
\begin{align*}
&  \int dx f^{\prime}(x)\mathcal{W}(\tau(x),R(x),f(x))\\
&  =\lim_{\s_{j}\rightarrow0}\sum_{j}\s_{j}\frac{f_{j+1}-f_{j}}{\s_{j}%
}\mathcal{W}_{j}(\tau_{j},R_{j},f_{j}).
\end{align*}
The functional $\Psi$ in (\ref{ansatz}) is then a product state over the
individual shell wave-functions and can be expressed as \begin{widetext}
\begin{equation}
\Psi[\tau,R,f] = \prod_j \Psi_j(\tau_j,R_j,f_j) = \prod_j \exp\left[-\frac{ic}{2l_p}
(f_{j+1}-f_j)[\tau_j + \cU_j(R_j,f_j)]\right].
\label{pprod}
\end{equation}
\end{widetext}The energy, $\varepsilon_{j}$, contained between lattice points
$j$ and $j+1$ is related to $f_{j+1}-f_{j}$ by
\begin{equation}
f_{j+1}-f_{j}=\frac{2G}{l_p c^{4}}(M_{j+1}-M_{j})c^{2}=\frac{2G}{l_pc^{4}}
\varepsilon_{j}\equiv\frac{2l_{p}}{c}\omega_{j},
\end{equation}
implying that the wave-functional can be expressed in the form
\begin{equation}
\Psi=\prod_{j}e^{-i\omega_{j}(\tau_{j}+\mathcal{U}_{j}(R_{j},F_{j}))}%
\end{equation}
where we have defined $\varepsilon_j = \hbar \omega_j = (M_{j+1}-M_j)c^2$. Our
strategy is to solve the Wheeler-DeWitt equation (\ref{wd}) independently at
each label, $j$. The continuum limit would be defined by putting together the
shell wave-functions and taking the limit as the lattice spacing approaches
zero. We shall see that the wave functions are independent of the original
choice of cell size, as it should be.

Before proceeding further it is necessary to define what is meant by a
functional derivative when functions are defined on a lattice. The defining
equations can be understood by analogy with the simplest properties of
functional derivatives of the functions $J(x).$%
\begin{eqnarray}
\frac{\delta J(y)}{\delta J(x)}  &=& \delta(y-x),\cr\cr
\frac{\delta}{\delta J(x)}\int dyJ(y)  &=& 1
\end{eqnarray}
From these definitions follows
\begin{equation}
\frac{\delta}{\delta J(x)}\int dyJ(y)\phi(y)=\phi(x).
\end{equation}
On a lattice we define, for the lattice intervals $x_{i}$ and $x_{j}$,%
\begin{equation}
\frac{\delta J(x_{i})}{\delta J(x_{j})}=\Delta(x_i-x_j) = \underset{\s_{i}
\rightarrow 0,~\forall i}{\lim }\frac{\delta_{ij}}{\s_{i}}\
\end{equation}
where $x_{i}$ labels the $i^{th}$ lattice site and $\delta_{ij}$ is just the
Kronecker $\delta$, equal to zero when the lattice sites $x_{i}$ and $x_{j}$
are different and one when they are the same. $\s_{i}$ is the size of the $i^{th}$
site, which we take to be arbitrarily given. Just as $\delta(y-x)$ is only
defined as an integrand in an integral, so $\Delta(x_{i}-x_{j})$ should also
be considered defined only as a summand in a sum over lattice sites. Hence%
\begin{equation}
\lim_{\s_i\rightarrow 0,~ \forall i} \frac{\delta}{\delta J_j(x_j)} \sum_i \s_i
J(x_i) = \underset{\s_{i}\rightarrow 0,~ \forall i}{\lim}%
{\displaystyle\sum\limits_{i}}
\s_{i}\frac{\delta J(x_{i})}{\delta J(x_{j})}=1
\end{equation}
and%
\begin{widetext}
\begin{equation}
\frac{\delta}{\delta J(x_j)} \sum_i \s_i J(x_i)\phi(x_i) = \lim_{\s_i\rightarrow 0,~
\forall i}\sum_i \s_i \Delta(x_i-x_j)\phi(x_i) = \phi(x_j)
\end{equation}
\end{widetext}
It follows that
\begin{equation}
\frac{\delta}{\delta J(x_{j})}\rightarrow \underset{\s_{j}\rightarrow 0,~\forall j}{\lim}~
\frac{1}{\s_{j}}\frac{\partial}{\partial J_j}
\end{equation}\\
where $J_j=J(x_j)$. This is compatible with the formal (continuum) definition of
functional derivative. With this regularization, the Wheeler-DeWitt equation on the
lattice may now be written for each lattice site, $j$, as%
\begin{widetext}
\begin{equation}
\underset{\s_{j}\rightarrow 0}{\lim}\left( \frac{1}{\s_{j}^{2}c^{2}}%
\frac{\partial^{2}}{\partial\tau_{j}^{2}}\pm\frac{1}{\s_{j}^{2}}\frac
{\partial^{2}}{\partial R_{\ast j}^{2}}\pm\frac{(f_{j+1}-f_{j})^{2}}{\s_{j}%
^{2}4l_{p}^{2}\left\vert \mathcal{F}_{j}\right\vert }\right)  \Psi_{j}%
(\tau_j,R_j,f_j)=0
\end{equation}
\end{widetext}
Factoring out the term $1/\s_{j}^{2}$ the equation becomes independent of the
choice of cell sizes,%
\begin{equation}
\left(  \frac{1}{c^{2}}\frac{\partial^{2}}{\partial\tau_{j}^{2}}\pm
\frac{\partial^{2}}{\partial R_{\ast j}^{2}}\pm\frac{(f_{j+1}-f_{j})^{2}}%
{4l_{p}^{2}\left\vert \mathcal{F}_{j}\right\vert }\right)  \Psi_{j}%
(\tau,R,F)=0.
\end{equation}
We see that the limit of vanishing cell size may be taken and that the
wave-functional is invariant under diffeomorphisms in this limit by
construction. Upon making some obvious replacements,%
\begin{equation}
\ \left[  \frac{\partial^{2}}{\partial R_{\ast j}^{2}}\mp\frac{\omega_{j}^{2}%
}{c^{2}}\left(  1\mp\frac{1}{\left\vert \mathcal{F}_{j}\right\vert }\right)
\right]  \Psi_{j}\left(  R_{j},F_{j}\right)=0,
\end{equation}
The upper sign refers to the exterior $(R>F)$ and the lower to the interior
$(R<F).$ $\Psi_{j}$ is the time independent part of the wave function.
Changing back from the coordinate $R_{\ast}$ to the original coordinate $R$,
the equation has the form%
\begin{equation}
\left\vert \mathcal{F}_{j}\right\vert \frac{\partial^{2}\Psi_{j}}{\partial
R_{j}^{2}}+\frac{1}{2}\frac{\partial\left\vert \mathcal{F}_{j}\right\vert
}{\partial R_{j}}\frac{\partial\Psi_{j}}{\partial R_{j}}\mp\frac{\omega
_{j}^{2}}{c^{2}}\left[  1\mp\frac{1}{\left\vert \mathcal{F}_{j}\right\vert
}\right]  \Psi_{j}=0
\end{equation}
where the upper sign refers to the exterior $(R>F)$ and the lower sign to the
interior $(R<F)$ and $\Psi_{j}$ refers to the time independent part of the
wave-functional in the $j^{th}$ cell. Defining the dimensionless variables
$z_{j}=R_{j}/F_{j}$ and $\gamma_{j}=F_{j}\omega_{j}/c$, redefining the time
independent part of the wave function $\Psi_{j}\equiv y_{j},$and suppressing
the subscript $j$ throughout, the equation for each shell takes the form
\begin{equation}
z(z-1)^{2}\frac{d^{2}y}{dz^{2}}+\frac{z-1}{2}\frac{dy}{dz}+\gamma^{2}%
z^{2}y = 0
\label{zeqn}%
\end{equation}
both in the exterior ($z>1$) and in the interior ($z<1$). \

\section{\label{sec:solutions}Solutions}

Before examining the solutions of (\ref{zeqn}) we recall some results of the
WKB approximation. The WKB solutions near spatial infinity were shown to be
\cite{vksw}%
\begin{equation}
\Psi_{\pm}^{\infty}\approx\exp\left[  \frac{-ic}{2l_{p}^{2}}%
{\displaystyle\int}drF^{\prime}(\tau\pm\frac{2F}{c}\sqrt{z})\right]
\end{equation}
In addition, finding the WKB approximation functional required no regularization.
This suggests that, for consistency, our shell wave-functions should agree with
the WKB approximation.

Writing (\ref{zeqn}) in terms of $u=1/z$ and keeping only the lowest order term in
$u$ we find
\begin{equation}
\frac{d^{2}y}{du^{2}}+\frac{2}{u}\frac{dy}{du}+\frac{\gamma^{2}}{u^{3}}y=0,
\end{equation}
which has as a solution
\begin{equation}
y_{\pm}(z)\approx A_{\pm}e^{\pm2i\gamma\sqrt{z}+\frac{1}{4}\ln z}.
\end{equation}
The time dependent wave-function then has the form
\begin{eqnarray}
\Psi_{\pm}^{\infty}  &=& e^{-i\sum_{j}\omega_{j}(\tau_{j}\pm\frac{2F_{j}}%
{c}\sqrt{z_{j}}+\frac{i}{4\omega_{j}}\ln z_{j})}\cr\cr
&=& \prod_{j}e^{-\frac{i\varepsilon_{j}}{\hbar}(\tau_{j}\pm\frac{2F_{j}}%
{c}\sqrt{z_{j}}+\frac{i\hbar}{4\varepsilon_{j}}\ln z_{j})}%
\label{asy}
\end{eqnarray}
We see that the last term is of order $\hbar$ and may be ignored when compared
with the WKB approximation. Hence the wave function we obtain at infinity
agrees with the results of the  WKB approximation at infinity.

For the asymptotic wave-functions of (\ref{asy}) to describe collapse, we must
choose shell wave-functions that correspond to incoming waves in the infinite
past and at large distances from the center. These are the wave-functions with
a positive sign in the exponent and they vanish along $\mathcal{I}^{+}$. The
time reversed situation, in which the wave-functions vanish on $\mathcal{I}%
^{-}$ but not on $\mathcal{I}^{+}$, are outgoing modes and represent expanding
matter. They are described by the wave-functions in (\ref{asy}) with the
negative sign in the exponent.

Near the horizon, $|z-1|\approx0$, we define $u=z-1$. Again, retaining only
the lowest orders in $u$, we find the effective equation
\begin{equation}
\frac{d^{2}y}{du^{2}}+\frac{1}{2u}\frac{dy}{du}+\frac{\gamma^{2}}{u^{2}%
}y\approx0\label{nearhor}%
\end{equation}
and the two independent solutions
\begin{equation}
\Psi^{\mathrm{hor}}\approx A_{\pm}^{\mathrm{hor}}\prod_{j}e^{-i\omega_{j}%
\tau_{j}+\frac{1}{4}(1\pm\sigma_{j})\ln\left\vert z_{j}-1\right\vert}
\end{equation}
where $\sigma_{j}=\sqrt{1-16\gamma_{j}^{2}}$. We consider only astrophysical
black holes, for which $M$ is large. Let $\gamma_{j}\gg1$, so that $\sigma
_{j}\approx 4i\gamma_{j}$, then the solution may be put in the form
\begin{equation}
{\Psi^{\mathrm{hor}}}\approx A_{\pm}^{\mathrm{hor}}\prod_{j}e^{-\frac
{i\varepsilon_{j}}{\hbar}(\tau_{j}\pm\frac{F_{j}}{c}\ln\left\vert
z_{j}-1\right\vert +\frac{i\hbar}{4\varepsilon_{j}}\ln\left\vert
z_{j}-1\right\vert )}%
\end{equation}
As before, the last term in the exponent above is small compared with the
first two and may be neglected to leading order. Then we recover the
near-horizon WKB solutions of \cite{vksw}, where it was shown that the
logarithmic term, which represents the scattering of the wave-function in the
background geometry, yields Hawking radiation at the Hawking temperature. In
the infinite future, the logarithmic term approaches negative infinity in the
approach to the horizon, therefore on $\mathcal{I}^{+}$ only the component
with the positive sign is relevant and we have
\begin{equation}
{\Psi^{\mathrm{hor}}}\approx A_{\pm}^{\mathrm{hor}}\prod_{j}e^{-\frac
{i\varepsilon_{j}}{\hbar}(\tau_{j}+\frac{F_{j}}{c}\ln\left\vert z_{j}%
-1\right\vert +\frac{i\hbar}{4\varepsilon_{j}}\ln\left\vert z_{j}-1\right\vert
)}%
\label{horizon0}%
\end{equation}
representing outgoing shells, scattered near their horizons.

Exact solutions to (\ref{zeqn}) may be obtained in a neighborhood of ordinary
points or about non-essentially singular points by expanding in a Frobenius
series. From the point of view of Hawking radiation, the horizon is the more
interesting of the two non-essential singularities of (\ref{zeqn}). Expanding
about the horizon will yield at least one solution by Fuch's theorem. Instead
of $z$ then, it is convenient to use the variable $u=z-1$. Assuming a solution
of the form
\begin{equation}
y(u)={\sum_{n=0}^{\infty}}a_{n}u^{k+n}\label{series}%
\end{equation}
we find that the roots
\begin{equation}
k_{\pm}=\frac{1}{4}\left(  1\pm\sqrt{1-16\gamma^{2}}\right)  \label{roo}%
\end{equation}
of the indicial equation
\begin{equation}
k\left(  k-\frac{1}{2}\right)  +\gamma^{2}=0
\end{equation}
differ by a non-integral number and therefore two independent solutions are
obtained, each defined by the following recursion relations,
\begin{widetext}
\begin{eqnarray}
&&a_1^\pm = -\frac{[k_\pm(k_\pm-1)+2\gamma^2]a_0^\pm}{(k_\pm+1)(k_\pm+\frac{1}{2})+\gamma^2}\cr\cr
&&a_{n+2}^\pm = - \frac{\gamma^2 a_n^\pm + [(k_\pm+n+1)(k_\pm+n)+2\gamma^2]a_{n+1}^\pm}{\gamma^2
+ (k_\pm+n+2)(k_\pm+n+\frac{3}{2})}.
\end{eqnarray}
\end{widetext}Each series provides an exact solution for small $|z-1|$, and
can be shown to be convergent for $|u|=|z-1|<1$. The general solution at each
shell is therefore a linear combination of the two,
\begin{equation}
{\widetilde{\Psi}}_{j,\pm}[u_{j}]={\sum_{n=0}^{\infty}}a_{j,n}^{\pm}%
u_{j}^{k_{j,\pm}+n},
\end{equation}
where we have reinserted the shell label, $j$. The very first order in the
expansion gives precisely (\ref{horizon0}) and therefore Hawking's formula.
Because this is the exact quantum state of the LTB class of collapse models it
also contains all the information about the quantum evolution of the collapse
close to the horizon, up to the scale of validity of the canonical theory. We
have shown that the first order approximation yields the standard Hawking
radiation at the Hawking temperature \cite{vksw}. The higher orders represent
corrections to Hawking's original formula.

\section{\label{sec:hawking} Hawking Radiation}

We now demonstrate that the leading correction to Hawking radiation makes the
spectrum non-thermal. First we show that the Hawking spectrum itself arises
from retaining only the $n=0$ term and dropping all terms of $\mathcal{O}%
(\hbar)$ in the exponent of (\ref{asy}). If one further assumes that
$\gamma\gg1$, then (\ref{roo}) gives $k_{\pm}\approx1/4 \pm i\gamma$, and the
above shell by shell wave-functions can be written as (dropping the shell
labels)
\begin{equation}
{\Psi^{\mathrm{hor}}} \approx A^{\mathrm{hor}}\exp\left[  -i\omega(\tau+
\frac{F}{c} \ln\left|  z-1\right|  )\right]  \label{int1}%
\end{equation}
and the asymptotic WKB wave-functions in (\ref{asy}) are
\begin{equation}
\Psi^{\infty}_{\pm}\approx A_\pm \exp\left[  -i\omega\left(  \tau\pm\frac
{2F}{c} \sqrt{z}\right)  \right]  . \label{int2}%
\end{equation}
Taking the negative sign in the exponent, the function $\Psi^{\infty}_{-}$ in
(\ref{int2}) represents a free outgoing dust mode of frequency $\omega$. We
take a complete basis of outgoing modes to be given by
\begin{equation}
\Psi^{\infty}_{-} = \exp\left[  -i\omega^{\prime}\left(  \tau-\frac{2F}%
{c}\sqrt{z} \right)  \right],
\label{asyf}
\end{equation}
letting $\omega^{\prime}$ take all possible values between zero and infinity.

The states ${\Psi^{\mathrm{hor}}}$ are the scattered modes near the horizon.
We are interested in computing the projection of (\ref{int1}) on the negative
frequency modes of the outgoing basis. For this purpose, we must consider the
inner product of the states on a hypersurface of constant Schwarzschild time.
As explained in \cite{vksw}, the Bogoliubov $\beta$-coefficient of interest is
given by
\begin{equation}
\beta(\omega,\omega^{\prime})=\langle{\Psi_{\omega^{\prime}}%
^{\infty}}^{\dagger}|\Psi_{\omega}^{\mathrm{hor}}\rangle=F\int
_{1}^{\infty}\frac{zdz}{z-1}\Psi_{\omega^{\prime}}^{\infty}\Psi_{\omega
}^{\mathrm{hor}}.
\label{bogo}
\end{equation}
The measure in the above integral is determined by transforming to the
Schwarzschild Killing time, $T$, which is related to the Tolman-Bondi (proper)
time by a standard transformation \cite{vksw}. The wave-functionals considered
must also be expressed in terms of the Schwarzschild Killing time. Ignoring
the unimportant $T$-dependent part (which will go away on squaring) the
Bogoliubov coefficient turns out to be
\begin{equation}
\beta(\omega,\omega^{\prime})\approx F\int_{1}^{\infty}\frac{zdz}{z-1}%
e^{\frac{4iF\omega^{\prime}}{c}\sqrt{z}}(z-1)^{-2i\frac{F\omega}{c}%
}\label{bigg}%
\end{equation}
This integral can be performed, after making the substitution $s=\sqrt{z}-1$.
Retaining only the lowest order in $s$ one gets
\begin{equation}
\beta(\omega,\omega^{\prime})\approx2Fe^{\frac{4iF\omega^{\prime}}{c}}\int
_{0}^{\infty}dse^{\frac{4iF\omega^{\prime}}{c}s}[2s]^{-1-2i\frac{F\omega}{c}%
}.\label{buu}%
\end{equation}
After squaring, one obtains the desired transition probability as
\begin{equation}
|\beta(\omega,\omega^{\prime})|^{2}=2\pi^{2}F^{2}\frac{kT_{H}}{\varepsilon
}\frac{1}{e^{\frac{\varepsilon}{kT_{H}}}-1},\label{hawk}%
\end{equation}
which is the Hawking spectrum at the Hawking temperature, $kT_{H}=\hbar
c^{3}/8\pi GM$. Note that $M$ is the total mass contained within the radiating shell.

Apart from neglecting all $\mathcal{O}(\hbar)$ corrections to the
wave-functionals, the Hawking formula is recovered by employing a
\textquotedblleft near horizon\textquotedblright\ approximation in which only
the lowest order in $s=\sqrt{z}-1$ is considered. The latter is justified by
the fact that only the near horizon region contributes significantly to the
integral in (\ref{bigg}) through the pole at $z=1$.

Consider the correction to this formula, which is obtained by taking into account
the first $\mathcal{O}(\hbar)$ correction. Keeping the asymptotic form of the
wave-functional the same, as given in (\ref{asyf}), but including the first
$\mathcal{O}(\hbar)$ correction to the near horizon wave-functional in (\ref{horizon0}),
we evaluate the Bogoliubov coefficient in (\ref{bogo}) to leading approximation in $s$
({\it i.e.,} near the horizon). With $s=\sqrt{z}-1$, this is
\begin{equation}
\beta(\omega,\omega^{\prime})\approx2Fe^{\frac{4iF\omega^{\prime}}{c}}\int
_{0}^{\infty}dse^{\frac{4iF\omega^{\prime}}{c}s}[2s]^{-\frac{3}{4}%
-2i\frac{F\omega}{c}}%
\label{corrected}
\end{equation}
Although it appears that the r.h.s. of (\ref{corrected}) has the same structure as
the WKB expression, it is evident that $\vert\beta(\omega,\omega')\vert^2$ is no longer
thermal. When $\gamma\gg 1$ we find
\begin{widetext}
\begin{equation}
|\beta(\omega,\omega')|^2 \approx 2\pi^2 F^2 \sqrt{\frac{2c}{F\omega'}}
\frac{kT_H}{\varepsilon}\frac{1}{e^{\frac{\varepsilon}{kT_H}}-1} \left[1 +
\frac{1}{2} \mathrm{Re}~ \psi\left(-\frac{i\varepsilon}{2\pi kT_H}\right) \right]
\end{equation}
\end{widetext}where $\psi(z)$ is the Polygamma function. As the first
correction depends on the Hawking temperature (or the mass $M$) and not just
the emission frequency, it cannot be associated simply with a correction to
the density of states. Furthermore, it cannot be obtained by modifying the
Hawking temperature. When $\gamma\gg1$, the real part of the Polygamma
function can be approximated by $\mathrm{Re}[\psi(-iy)]\approx\ln y$ and we
find, \begin{widetext}
\begin{equation}
|\beta(\omega,\omega')|^2 \approx 2\pi^2 F^2\sqrt{\frac{2c}{F\omega'}} \frac{kT_H}
{\varepsilon} \frac{1} {e^{\frac{\varepsilon}{kT_H}}-1}\left[1 - \frac{1}{2}\ln
\left(\frac{\pi kT_H}{\varepsilon}
\right)\right],
\end{equation}
\end{widetext}which completes the first correction to the WKB approximation.
The correction term is of order $\mathcal{O}(\ln\hbar).$

\section{\label{sec:discussion} Discussion}

The correction we have obtained cannot be accounted for by modifying the Hawking temperature
and we conclude that it renders the radiation non-thermal. The breakdown of
Hawking's thermal spectrum, when quantum gravitational effects are accounted
for \textit{i.e.,} when we go beyond the WKB approximation, is not equivalent
to a correction to the black hole entropy. Corrections to the black hole entropy
formula have been computed in various approaches before \cite{mdc}, but they can
all be understood as relating to a thermal spectrum, simply modifying the
Hawking temperature. Put in another way, we have shown that there are
``grey body'' factors whose origin is quantum gravitational and they are
associated with the horizon. They are not to be confused with grey body factors
that are associated with the radiation at spatial infinity, which originate in
the backscattering of the radiation against the classical space-time geometry.

The thermal character of the Hawking radiation from collapsing matter is
sometimes invoked to suggest that quantum gravitational evolution may not be
unitary and that information is lost. Our result indicates that this is not
necessarily the case, at least within the context and limitations of the
canonical theory. A more complete picture will be obtained from a deeper
understanding of the behavior of solutions to (\ref{zeqn}). We hope to come
back to a more developed discussion of the properties of the solutions to this
equation in a future publication. One of the issues we expect to discuss will
be the nature of the singularity at infinity. It is an essential singularity
so there will be an issue regarding normalizing the wave function. The
collapsing system we study has been a marginally bound system - physically
this means that the total vacuum energy is zero. From quantum considerations
we expect there to be a non-vanishing zero point energy. A preliminary
estimate of the result of allowing a non-zero vacuum energy is that the
singularity at infinity will be softened but that the general features of the
solution discussed so far will be unchanged.

It was especially gratifying to see that the results are completely
independent of the choice of cell sizes in the lattice.

\section*{Acknowledgments}

We acknowledge the partial support of the Funda\c{c}\~ao para a Ci\^encia e a
Tecnologia (FCT), Portugal, under contract POCTI/32694/FIS/2000. L.W. was
supported in part by the Department of Energy, USA, under Contract Number DOE-FG02-84ER40153.


\vskip 5mm

\begin{thebibliography}{99}                                                                                               %


\bibitem {kt}H. A. Kastrup and T. Thiemann, Nucl. Phys. \textbf{B425}, (1994) 665.

\bibitem {kuc1}K. Kucha\v r and C. Torre, Phys. Rev. \textbf{D43}, (1991) 419.

\bibitem {kuc2}K. Kucha\v r, Phys. Rev. \textbf{D50}, (1994) 3961 .

\bibitem {kuc3}J. Brown and K. Kucha\v r, Phys. Rev. \textbf{D51} (1995) 5600.

\bibitem {vws}Cenalo Vaz, L. Witten and T.P. Singh, Phys. Rev. \textbf{D63} (2001) 104020.

\bibitem {ltb1}G. LeMa\^itre, Ann. Soc. Sci. Bruxelles I, \textbf{A53} (1933)
51;\newline R. Tolman, Proc. Natl. Acad. Sci. USA \textbf{20} (1934)
410;\newline H. Bondi, Mon. Not. Astron. Soc. \textbf{107} (1947) 343.

\bibitem {bek1}J.D. Bekenstein, Ph. D. Thesis, Princeton University, April
1972;\newline\textit{ibid}, Lett. Nuovo Cimento \textbf{4} (1972)
737;\newline\textit{ibid}, Phys. Rev. \textbf{D7} (1973) 2333; \textit{ibid},
Phys. Rev. \textbf{D9} (1974);\newline\textit{ibid}, Phys. Rev. Lett.
\textbf{70} (1993) 3680;\newline\textit{ibid}, Phys. Lett. \textbf{B360}
(1995) 7.

\bibitem {cvlw1}Cenalo Vaz and L. Witten, Phys. Rev. \textbf{D60} (1999)
024009;\newline Cenalo Vaz, Phys. Rev. \textbf{D61} (2000) 064017;\newline Cenalo Vaz
and L. Witten, Phys. Rev. \textbf{D63} (2001) 024008;\newline\textit{ibid},
Phys. Rev. \textbf{D64} (2001) 084005.

\bibitem {vksw}Cenalo Vaz, C. Kiefer, T. P. Singh and L. Witten, Phys. Rev.
\textbf{D67} (2003) 024014.

\bibitem {bk}J.-G. Demers and C. Kiefer, Phys. Rev. D53, 7050 (1996).\newline
T. Brotz and C. Kiefer, Phys. Rev. D55, 2186 (1997).

\bibitem {biku}J. Bi\v cak and K. Kucha\v r, Phys. Rev. \textbf{D56} (1997) 4878.

\bibitem {lwf}J. Louko, B. Whiting and J. Friedman, Phys. Rev. \textbf{D57}
(1998) 2279.

\bibitem {haki}P. H\'aj\'\i\v cek and C. Kiefer, Int. J. Mod. Phys.
\textbf{D10} (2001) 775;\newline\textit{ibid}, Nucl. Phys. \textbf{B603}
(2001) 531.

\bibitem {ha}P. H\'aj\'\i\v cek, Nucl. Phys. \textbf{B603} (2001) 555.

\bibitem {vws2}Cenalo Vaz, L. Witten and T.P. Singh, Phys. Rev. \textbf{D65}
(2002) 104016.

\bibitem {deW}B.S. DeWitt, Phys. Rev. \textbf{160} (1967) 1113.

\bibitem {mdc}D. Youm, Phys. Rept. \textbf{316} (1999) 1.\newline R.K. Kaul
and P. Majumdar, Phys. Rev. Lett. \textbf{84} (2000) 5255.
\end{thebibliography}
\end{document}